\documentclass[twocolumn,amsmath,amssymb,prb,footinbib,floatfix]{revtex4}
\usepackage{amssymb}
\usepackage{amsmath}
\usepackage{amsthm}
\usepackage{helvet}
\usepackage{hyperref}
\usepackage{dsfont}
\usepackage[usenames,dvipsnames]{color}

\newcommand{\id}{\mathds{1}}

\newcommand{\be}{\begin{eqnarray}}
\newcommand{\ee}{\end{eqnarray}}
\newcommand{\non}{\nonumber}

\makeindex

\begin{document}

\title{Free fermionic and parafermionic quantum spin chains with multispin interactions}
\date{\today}

\author{Francisco C. Alcaraz}
\email{alcaraz@ifsc.usp.br}
\affiliation{Instituto de F\'{\i}sica de S\~ao Carlos, Universidade de S\~ao
  Paulo, Caixa Postal 369, S\~ao Carlos, SP, Brazil}

\author{Rodrigo A. Pimenta}
\email{pimenta@ifsc.usp.br}
\affiliation{Instituto de F\'{\i}sica de S\~ao Carlos, Universidade de S\~ao
  Paulo, Caixa Postal 369, S\~ao Carlos, SP, Brazil}

\begin{abstract}
We introduce a new a family of $Z(N)$ multispins quantum chains
with a free-fermionic ($N=2$) or free-parafermionic ($N>2$) eigenspectrum. The models
have $(p+1)$ interacting spins ($p=1,2,\dots$), being Hermitian in the $Z(2)$ (Ising) case
and non-Hermitian for $N>2$. We construct a set of mutually commuting charges that allows us
to derive the eigenenergies in terms of the roots of polynomials generated by a recurrence relation of order $(p+1)$. In the critical limit
we identify these polynomials with certain hypergeometric polynomials ${}_{p+1}F_p$.
Also in the critical regime, we calculate the ground state energy
in the bulk limit and verify that they are given in terms of the Lauricella hypergeometric series.
The models with special couplings are self-dual and at the self-dual point show a critical behavior
with dynamical critical exponent $z_c=\frac{p+1}{N}$.
\end{abstract}


\maketitle

{\it\bf Introduction}. Models known by the general name of free systems play a crucial
role in condensed matter physics and in statistical mechanics
of exactly integrable systems.  The simplest and prototype of these
models is the quantum Ising model in a transverse field \cite{Schultz:1964fv,Pfeuty}.
The Hamiltonian defined in a chain of $L$ sites and free boundary conditions
is given by,
\be\label{HIsing}
\mathcal{H}_{I} =  - \sum_{i=1}^{L-1} \lambda_{2i-1}\sigma_i^z\sigma_{i+1}^z-\sum_{i=1}^L \lambda_{2i} \sigma_i^x\,,
\ee
where $\sigma_i^{x,z}$ are the standard spin-$\frac{1}{2}$ Pauli matrices
attached at the sites ($i=1,\dots,L$) and the couplings
$\lambda_i$ ($i=1,\dots,2L-1$) play the role of local temperatures.
The $2^L$ eigenvalues of this model can be obtained through
a Jordan-Wigner transformation and it has the form,
\be\label{EIsing}
E_{I} = \pm \epsilon_1 \pm \epsilon_2 \pm \cdots \pm \epsilon_L\,,
\ee
where the pseudo energies $\epsilon_i$ ($i=1,\dots,L$), which are functions of
$\{\lambda_i\}$. The model is called free since all the $2^L$
values of $E_{I}$ are obtained solely from the $L$ pseudo-energies $\epsilon_i$. In this case we have a free-fermionic
system since each pseudo-energy can appear with both signals
as in a standard fermionic system.

A simple direct generalization of (\ref{HIsing}) with $Z(N)$ symmetry and
having a free eigenspectrum was proposed in the late 80's \cite{Baxter:1989bma,Baxter:1989vv}:
\be\label{HBaxter}
\mathcal{H}_{B} =  - \sum_{i=1}^{L-1} \lambda_{2i-1}Z_i^\dagger Z_{i+1}-\sum_{i=1}^L \lambda_{2i}X_i\,,
\ee
where $X_i$ and $Z_i$ ($i=1,\dots,L$) are obtained from the $Z(N)$ generalizations of the Pauli matrices satisfying the algebra,
\be\label{Weyl}
ZX = \omega XZ\,, ~ X^N = Z^N = 1\,, ~ Z^\dagger = Z^{N-1}\,,~ \omega = e^{\frac{2i\pi}{N}}\,,
\ee
and having the unique irreducible representation \cite{Weyl} in $\mathbb{C}^N$ given by,
\be
X_{j,k}=\delta _{j,k+1} \quad (\textrm{mod}~N)\,,\quad
Z_{j,k}=\omega^{j-1} \delta _{j,k}\,.
\ee
The spectrum of (\ref{HBaxter}) has the
form,
\be\label{EBaxter}
-E_B = \omega^{s_1} \epsilon_1 + \omega^{s_2} \epsilon_2 + \cdots + \omega^{s_L} \epsilon_L
\ee
where $s_i \in \{0,1,\dots,N-1\}$ and $\epsilon_i$ are given in terms of the roots of a given polynomial
\cite{Baxter:1989bma,Baxter:1989vv}.
Clearly, the result (\ref{EBaxter}) reduces to (\ref{EIsing}) when $N=2$. The formula (\ref{EBaxter}) was conjectured in \cite{Baxter:1989bma,Baxter:1989vv} and proved
in \cite{Tarasov:1991mf,Baxfunc}. More recently, it was proved using raising and lowering parafermionic operators \cite{Fendley:2013snq}, generalizing the Clifford algebra method \cite{Kaufman:1949ks} for the free parafermionic system. The work \cite{Fendley:2013snq}
has boosted a number of papers on the spectral problem of (\ref{HBaxter}) 
\cite{Baxter2014,
Au_Yang_2014,auyang2016parafermions,Alcaraz_2017,Alcaraz_2018,Liu_2019}. In
analogy with the Ising chain, the model (\ref{HBaxter}) is free in the sense that
the quasi-energies are independent of the choice of $s_i$, despite the model is not Hermitian
for $N>2$.

We recall that the Hamiltonian (\ref{HBaxter}) can be derived from the
transfer matrix of the so-called $\tau_2$ model \cite{Bazhanov:1989nc,Baxter:1999mn}
under certain restrictions. For the general Hamiltonian arising from
the $\tau_2$ model see \textit{e.g.} \cite{Tarasov:1991mf,Baxter2014}.
To the best our knowledge, the Hamiltonian (\ref{HBaxter})
or the general case derived from the $\tau_2$ model are so far
the only known models with a spectrum of the form (\ref{EBaxter}) when $N>2$.

The solution of the Ising chain involves the transformation of the Pauli operators in
fermions by means of the Jordan-Wigner transformation. As a result, the Hamiltonian
acquires a quadratic form in Fermi or Majorana operators. Similarly, using the
Fradkin-Kadanoff transformation \cite{Fradkin:1980th}, the Hamiltonian (\ref{HBaxter}) can be written in terms
of parafermions \cite{Fendley:2012vv,Fendley:2013snq}. However, the resulting Hamiltonian is not
quadratic in the parafermions, and yet it has a free parafermionic spectrum.

Recently, a remarkably simple Hamiltonian with 3-spin interactions that has a spectrum of the form (\ref{EIsing})
was introduced \cite{Fendley:2019sdx}. It is given by,
\be\label{HFendley}
\mathcal{H}_F = -\sum_{i=1}^{L-2} \lambda_{i}\sigma_i^z\sigma_{i+1}^x\sigma_{i+2}^x\,.
\ee
The standard Jordan-Wigner transformation applied to (\ref{HFendley}) yields
a Hamiltonian of order four in the Majorana fermions. Given this feature, it is quite surprising that it has a free spectrum.

The common feature of the Hamiltonians (\ref{HIsing},\ref{HBaxter},\ref{HFendley}) is that
its energy density operators satisfy a very simple algebra \cite{Fendley:2013snq,Fendley:2019sdx} with $M$ generators $h_a$, namely,
\be\label{halgebra}
h_a h_{a+m} &=& \omega h_{a+m}h_a \quad \textrm{for}\quad 1\leq	m \leq p\,,\non\\
\left[h_a, h_b\right] &=& 0 \quad \textrm{for} \quad |a-b|>p \,,\quad
h_a^N = \lambda_a^N\,,
\ee
where $p=1$ for (\ref{HIsing},\ref{HBaxter}) and $N=p=2$ for (\ref{HFendley}). In
fact, the Hamiltonians (\ref{HIsing},\ref{HBaxter},\ref{HFendley}) can be
written as,
\be\label{Hgen}
-\mathcal{H} = \sum_{i=1}^M h_i\,,
\ee
where $M=2L-1$,
\be
h_{2i-1} &=& \lambda_{2i}X_i\quad \textrm{for}\quad i =1,\dots,L\,,\non\\
h_{2i} &=&  \lambda_{2i-1}Z_iZ_{i+1}^\dagger\,\quad \textrm{for}\quad i =1,\dots,L-1 \,,
\ee
for (\ref{HIsing},\ref{HBaxter}) and $M=L-2$,
\be
h_{i} &=& \lambda_{i}\sigma_i^z\sigma_{i+1}^x\sigma_{i+2}^x\quad \textrm{for}\quad i =1,\dots,L-2\,,
\ee
for (\ref{HFendley}).

The algebra (\ref{halgebra}) for $p=1$ belongs to the realm of generalized Clifford algebras, see \textit{e.g.}
\cite{Morris1967,Morris1968,Mittag,Martin1989,Truong1986,K1986,Ramakrishnan1986,Jaffe:2014ama}.

The aim of this letter is to argue that Hamiltonians of the form (\ref{Hgen}) whose
energy density operators satisfy (\ref{halgebra}) for arbitrary positive integer $p$ and $N$ 
have a free parafermionic spectrum of the form (\ref{EBaxter}).
In order to that, we follow the strategy in \cite{Fendley:2019sdx}. The idea is that,
using (\ref{halgebra}), one can easily build a set of conserved charges
associated with (\ref{Hgen}) and use them to construct a generating function.
This generating function satisfies an important product formula  (\ref{prodformula}),
which is similar to that satisfied by the transfer matrix of the $\tau_2$ model \cite{Baxfunc}.
We remark that this procedure is independent of the representation of the algebra (\ref{halgebra}). Nevertheless,
it is easy to find representations of (\ref{halgebra}) using (\ref{Weyl}) which leads to interesting quantum
spin chains with $(p+1)$ multispin interactions. For example, 
$h_i= \lambda_i Z_i Z_{i+1}\cdots Z_{i+p-1} X_{i+p}$ 
produces a Hamiltonian acting in the vector space $\left(\mathbb{C}^{N}\right)^{\otimes {M+p}}$:
\be\label{H}
\mathcal{H} = -\sum_{i=1}^M h_i = -\sum_{i=1}^M \lambda_i Z_i Z_{i+1}\cdots Z_{i+p-1} X_{i+p}.
\ee

{\it\bf The commuting charges and its generating function}.
The simple form of the algebra (\ref{halgebra}) allows one to construct
explicitly
a set of mutually commuting charges, the Hamiltonian being one of them. 
As in \cite{Fendley:2019sdx} for the case $p=2$, the commuting charge operators $H_M^{(l)}$
are obtained by summing all the products $h_{j_1}h_{j_2}\cdots h_{j_l}$
formed by $l$ generators $h_i$ in the set $\{h_1,\dots,h_M\}$ whose indices difference
$|j_i-j_{i\pm 1}|$ is larger then $(p+1)$, namely:
{\allowdisplaybreaks
\be\label{charges}
H_M^{(0)} &=& \id\,,\quad
H_M^{(1)} = -H = \sum_{j=1}^Mh_j\,,\non\\
H_M^{(2)} &=& \sum_{j_1=1}^M\sum_{j_2=j_1+p}^M h_{j_1}h_{j_2} \,,\quad\cdots\quad,
\\
H_M^{({\bar M})} &=& \sum_{j_1=1}^M\sum_{j_2=j_1+p+1}^M
\cdots \sum_{j_{\bar M}=j_{{\bar M}-1}+p+1}^M h_{j_1}h_{j_2}\dots h_{j_{\bar M}}\,,\non
\ee}
where ${\bar M}$ is the largest number of commuting $h_j$
we can obtain from the set $\{h_1,\dots,h_M\}$, and it is given by the 
integer part of $\frac{M+p}{M+1}$, \textit{i.e.}, ${\bar M}=\lfloor \frac{M+p}{p+1}\rfloor$.
We notice that the Hamiltonian (\ref{Hgen})
is the charge $-H_M^{(1)}$.

The generating function of the charges is defined as,
\be\label{transfer}
G_M(u) = \sum_{l=0}^{\bar M}(-u)^l H_M^{(l)}\,,
\ee
and we claim that it satisfies the fundamental
properties,
\be\label{commute}
[G_M(u),G_M(v)] =[H_M^{(l)},G_M(u)]=[H_M^{(l)},H_M^{(l')}] = 0,
\ee
for arbitrary $u$ and $v$ and $l,l'=0,1,\dots,\bar M$.

Actually, the commutativity relations (\ref{commute}) follow from the fact the generating function
$G_M(u)$ and the charges $H_M^{({l})}$ satisfy recurrence relations,
namely,
\be\label{recurrencecharges}
H_M^{(l)} = H_{M-1}^{(l)}+h_M H_{M-(p+1)}^{(l-1)}\,,
\ee
\be\label{recurrenceformula}
G_M(u) = G_{M-1}(u)-uh_M G_{M-(p+1)}(u)\,,
\ee
with the initial conditions $H_M^{(0)}=\id$, $H_M^{(l)}=0$ for $l<0$ or $M\leq 0$, and $G_M(u)=\id$ for $M\leq 0$.
The proof \cite{AP2020} follows from the fact that,
\be
\beta_M^{(l)}(u) = \left[H_M^{(l)},G_M(u)\right]\,,
\ee
satisfies a recurrence relation with the same form as (\ref{recurrenceformula}), namely,
\be\label{betarec}
\beta_M^{(l)}(u) = \beta_{M-1}^{(l)}(u) -uh_M \beta_{M-(p+1)}^{(l)}(u)\,.
\ee
We remark that (\ref{commute}) has been proved for $N=p=2$ in \cite{Fendley:2019sdx}
using the fact the generating function admits certain factorizations and it was also  proved for
$p=1,N=3$ in \cite{Fendley:2013snq}.

{\it\bf Product formula for the generating function}.
The free spectrum nature for the Hamiltonian (\ref{Hgen})
emerges
from the fact that generating function $G_M(u)$ satisfy an important product:
\be\label{prodformula}
\tau_M(u)\equiv G_M(u)G_M(\omega u)\dots G_M(\omega^{N-1}u) = P_M^{(p)}(u^N)\id\,,\non\\
\ee
where $P_M^{(p)}(z)$ is a polynomial of degree $\bar M$ in $z=u^N$. The formula
(\ref{prodformula}) has the same form of the one satisfied
by the transfer matrix of the $\tau_2$ model \cite{Baxfunc}. When $N=2$, the formula
(\ref{prodformula}) is known as the inversion relation \cite{Baxbook}, since $G_M(-u)$ is
proportional to the inverse of $G_M(u)$.

The proof of (\ref{prodformula}) follows the same lines of the proof of (\ref{commute}).
The key point \cite{AP2020} is to show that $\tau_M(u)$ also satisfy a recurrence relation, namely,
\be\label{rectau}
\tau_M(u)=\tau_{M-1}(u)-u^Nh_M^N\tau_{M-(p+1)}(u)\,,
\ee
with $\tau_M(u)=1$ for $M\leq 0$. This expression gives $\tau_1(u)=1-u^Nh_1^N$ and by iterating (\ref{rectau})
we obtain that $\tau_M(u)$ is a polynomial in $u^N$, as advanced in (\ref{prodformula}). We see from (\ref{rectau})
that the coefficients of $P_M^{(p)}(u^N)$ depend on the values $h_i^N=\lambda_i^N$ ($i=1,\dots,M$) that are
given by the couplings $\lambda_i$ ($i=1,\dots,M$) defining the Hamiltonian in (\ref{Hgen}) and (\ref{H}).
The proof of (\ref{rectau}) also to rely on the recurrence equations (\ref{recurrencecharges},\ref{recurrenceformula}),
as we have already verified for the case $N=2$ and arbitrary $p$, and when $N=3$ for $p=1,2$. For other values of $N>2$, and several values of $p$, we verified
(\ref{prodformula}) numerically for small values of $\bar M$.
Based on our analytical and numerical analyses we conjecture that (\ref{prodformula}) is valid for arbitrary $N$ and $p$.
We remark that the case $N=p=2$ has been proved in \cite{Fendley:2019sdx} using the above mentioned
factorization of the transfer matrix.

The relation (\ref{rectau}) implies the recurrence relation for the polynomial (hereafter we use the variable $z=u^N$ for the polynomial $P_M^{(p)}$),
\be\label{recpol}
P_M^{(p)}(z)=P_{M-1}^{(p)}(z)-u^N\lambda_M^NP_{M-(p+1)}^{(p)}(z)\,,
\ee
with the initial condition $P_l^{(p)}(u^N)=1$ for $l\leq 0$. Comparing (\ref{recpol}) with
the recursion relation (\ref{recurrenceformula}) we identify
the dependence in $\lambda_i^N$ ($i=1,\dots,M$) of the coefficients $C_M^{(l)}$ in the expansion,
\be\label{generalP}
P_M^{(p)}(z)=\sum_{l=0}^{\bar M}(-z)^l C_M^{(l)}\,.
\ee
They are obtained by replacing $h_j\rightarrow \lambda_j^N$ in (\ref{charges}), \textit{i.e.},
\be\label{polcoef}
C_M^{(l)} = \sum_{j_1=1}^M\sum_{j_2=j_1+p+1}^M
\cdots \sum_{j_{\bar M}=j_{{\bar M}-1}+p+1}^M \lambda_{j_1}^N\lambda_{j_2}^N\dots \lambda_{j_l}^N \,,
\ee
for $l=0,1,\dots,\bar M$.

In the special case where $\lambda_j^N=1$ ($j=1,\dots,M$), $C_M^{(l)}$
corresponds to the number of ways to put $l$ particles with excluded
volume of $(p+1)$ lattice units in a lattice of size $M$, that is, it is given by the binomial coefficient,
\be
C_M^{(l)}=\binom{M-p(l-1)}{l}=\frac{(M-p(l-1))!}{(M-p(l-1)-l)!l!}\,,
\ee
and we can identify $P_M^{(p)}(z)$ as the generalized hypergeometric polynomial
${}_{p+1}F_p$ \cite{AeqB},
\begin{widetext}
\be\label{polP}
P_M^{(p)}(z) = {}_{p+1}F_p \left(\begin{matrix} -\frac{M+p}{p+1} \quad -\frac{M+p-1}{p+1} \quad -\frac{M+p-2}{p+1} \quad \cdots \quad -\frac{M}{p+1}\\
-\frac{M+p}{p} \quad -\frac{M+p-1}{p} \quad \cdots \quad -\frac{M+1}{p}\end{matrix}\,; \frac{(p+1)^{p+1}}{p^p}z\right)
= \sum_{l=0}^{\bar M}(-1)^l\binom{M-p(1-l)}{l}z^{l}\,.
\ee
\end{widetext}
It is worth noticing that for the critical quantum Ising chain, where $p=1$, the
polynomial (\ref{polP}) is related to the Chebyshev polynomial of second
type, \textit{i.e.}, $P_M^{(1)}(z) = z^{\frac{M+1}{2}}U_{M+1}\left(\frac{1}{2z^\frac{1}{2}}\right)$.

{\it\bf The solution of the product formula and the spectrum of the Hamiltonian}.
The eigenspectrum of the Hamiltonian (\ref{Hgen}) $-H_M^{(1)}$ as well as of all the charges
$H_M^{(l)}$ are obtained from the product formula (\ref{prodformula}). Thanks to (\ref{commute}),
by applying (\ref{prodformula}) to a given eigenfunction of $G_M(u)$ with
eigenvalue $\Lambda(u)$, we obtain,
\be\label{prodlam}
\Lambda(u)\dots\Lambda(\omega^{N-1}u)=P_M^{(p)}(u^N)=\prod_{i=1}^{\bar M}\left(1-\frac{u^N}{z_i}\right)\,,
\ee
where $z_i$ are the roots of (\ref{generalP}). The relation
(\ref{prodlam}) can be solved in terms of $z_i$,
\be\label{Lam}
\Lambda_M(u) = \prod_{i=1}^{\bar M} \left(1- u\frac{\omega^{s_i}}{z_i^{1/N}}\right)
= \prod_{i=1}^{\bar M} \left(1- u\,\omega^{s_i}\epsilon_i\right)\,,
\ee
where $\epsilon_i=z_i^{-1/N}$ and $s_i \in \{0,1,\dots,N-1\}$. We verified that
all the roots of $z_i$ of (\ref{generalP}) are distinct. Therefore, there are $N^{\bar M}$
distinct eigenvalues of the generating function. We can expand (\ref{Lam}) in terms of its roots,
\be\label{lamsol}
\Lambda_M^{\{s_i\}}(u) =
\sum_{l=0}^{\bar M} (-1)^le_l\left(\omega^{s_1}\epsilon_1,\dots,\omega^{s_{\bar M}}\epsilon_{\bar M}\right)u^l\,,
\ee
where $e_l\left(\omega^{s_1}\epsilon_1,\dots,\omega^{s_{\bar M}}\epsilon_{\bar M}\right)$
is the $l-$th elementary symmetric polynomial in
\allowdisplaybreaks{$\left\{\omega^{s_1}\epsilon_1,\dots,\omega^{s_{\bar M}}\epsilon_{\bar M}\right\}$}
\footnote{The $k-$th elementary symmetric polynomial in $n$ variables $\{x_1,x_2,\dots,x_n\}$ is defined by,
$
e_{k}(x_1,x_2,\dots,x_n)=\sum_{1\leq j_1 < j_2 < \cdots < j_k \leq n }  x_{j_1}x_{j_2}\cdots x_{j_k}
$ for $k=0,1,\dots,n$.}.

Since $u$ is arbitrary, applying (\ref{transfer}) to a eigenvector with eigenvalue $\Lambda_M^{\{s_i\}}(u)$
yields the eigenvalues $q_{\{s_i\}}^{(l)}$ of the charges $H_M^{(l)}$, \textit{i.e.},
\be
q_{\{s_i\}}^{(l)}=e_l\left(\omega^{s_1}\epsilon_1,\dots,\omega^{s_{\bar M}}\epsilon_{\bar M}\right),\quad l=1,\dots,\bar M\,.
\ee
In particular the Hamiltonian (\ref{Hgen}) has the the spectrum with the same form as (\ref{EBaxter}),
\be\label{EHgen}
-E^{\{s_i\}} = q_{\{s_i\}}^{(1)}= \omega^{s_1} \epsilon_1 + \omega^{s_2} \epsilon_2 + \cdots + \omega^{s_L} \epsilon_{\bar M}\,.
\ee

The spectrum of (\ref{Hgen}) is thus determined by the roots of the polynomial (\ref{generalP})\footnote{Note that all the levels may be degenerate depending
on the representation of the $h_i$ in (\ref{Hgen}). For the representation (\ref{H}) we can in fact find \cite{AP2020} several additional charges in involution with $H_M^{(l)}$ ($l=1,\dots,M$).}.
For
general values of the couplings $\lambda_i$ ($i=1,\dots,M$) we expect a quite rich phase diagram,
see \cite{Fendley:2019sdx} for the case $N=p=2$.

{\it\bf Criticality}. Instead of considering the most general models we are going to
restrict ourselves to the case where we have only two values
for the coupling constants (1 and $\lambda$). We consider the model where we alternate
the block of $p$ couplings with the values (1 and $\lambda$), \textit{i.e.},
\be\label{speccouplings}
\lambda_{kp+j} = \Bigg\{\begin{array}{l}
1,\quad k~~odd\\
\lambda,\quad k~~even
\end{array}\,,
\ee
for $k=0,1,2,\dots$ and $j=1,\dots,p$.

For $p=1$ we recover (\ref{HBaxter}) with the couplings $\lambda_{2i-1}=1,\lambda_{2i}=\lambda$, and for example
for $p=3$ we have,
\be
-\mathcal{H}(\lambda) &=& h_1+h_2+h_3 + \lambda (h_4+h_5+h_6)
\non\\&+&h_7+h_8+h_9 + \cdots\,.
\ee
This model is gapped (non-critical) at the limiting values $\lambda=0$ and $\lambda\rightarrow\infty$,
since the model reduces to a set of disjoint $p-$interacting systems. On the other hand by
shifting the variables $h_i\rightarrow h_{i+p}$ we see that apart from the boundary conditions the eigenenergies
of the Hamiltonians with couplings $\lambda$ and $1/\lambda$ are related, \textit{i.e.},
\be
\mathcal{H}(\lambda)=\frac{1}{\lambda}\mathcal{H}\left(\frac{1}{\lambda}\right)\,,
\ee
and thus the model is self-dual, and in the case we have a single transition
from $\lambda=0$ and $\lambda\rightarrow \infty$, it is the phase transition point $\lambda=\lambda_c=1$. This is the case for $p=1$ \cite{Alcaraz_2017} and
$N=p=2$ \cite{Fendley:2019sdx}, and for general values of $p$,
as we argue below.

In order to show the criticality at $\lambda=\lambda_c=1$ we extend the interesting procedure
introduced in \cite{Fendley:2019sdx} for the case $p=2$, that enable us to find the roots of (\ref{polP}) in the
bulk limit $M\rightarrow\infty$.

We consider lattice sizes that are multiples of $(p+1)$, \textit{i.e.}, $M=(p+1)\bar M$. In this case,
the recurrence relation
(\ref{recpol}) for the polynomials (\ref{polP}) gives,
\be\label{multirec}
P_M^{(p)}(z) = P_{M-(p+1)}^{(p)}(z) - \sum_{i=1}^{p+1} \binom{p+1}{i} z^{i} P_{M-i(p+1)}^{(p)}(z),\non
\ee
with the condition $P_{(p+1)l}^{(p)}(z)=0$ for $l\leq 0$.

If we have a root $z_k$ of the $M=(p+1)\bar M$-polynomial,
\textit{i.e.}, $P_{(p+1)\bar M}^{(p)}(z_k)=0$, this root should satisfy the
$\bar M+1$ coupled difference equations of order $p+1$,
\be\label{difeqP}
P_{(p+1)(\bar M-j)}^{(p)}(z_k)=\sum_{i=0}^{p+1}\binom{p+1}{i}z_k^{i}P_{(p+1)(\bar M-j-(i-1))}^{(p)}(z_k)\,,
\ee
with $j=1,\dots,{\bar M}$ and initial conditions,
\be\label{initdifeqP}
P_{(p+1)\bar M}^{(p)}(z_k)=P_{(p+1)l}^{(p)}(z_k)=0\,,\quad l=1,\dots,p\,.
\ee
The ansatz
\be\label{ansaP}
P_{(p+1)(\bar M-j)}^{(p)}(z_k) = \mu^{\bar M+p-1-j}z_k^{1-j}\,,~j=1,\dots,\bar M,
\ee
gives the characteristic equation,
\be\label{polQ}
(\mu+1)^{p+1}-z_k^{-1} \mu^p=(\mu-\mu_1)\cdots (\mu-\mu_{p+1})=0\,.
\ee
The general solution of (\ref{difeqP}) is then given by the combination of the
$(p+1)$ roots $\mu_i$ of the characteristic equation,
\be
P_{(p+1)(\bar M-j)}^{(p)}(z_k)=\sum_{i=1}^{p+1}A_iz_k^{1-j} \mu_i^{\bar M+p-1-j}\,,
\ee
where $A_i$ ($i=1,\dots,p+1$) are going to be fixed by (\ref{initdifeqP}), \textit{i.e.},
\be\label{eqAi}
\sum_{i=1}^{p+1}A_i \mu_i^{\bar M+p-1} = 0, ~ \sum_{i=1}^{p+1}A_i \mu_i^{p-l-1} = 0,~~ l=1,\dots,p,\,
\ee
where we set $A_{p+1}=1$. 

We could not find a closed form for the roots of the polynomial (\ref{polQ}). However, surprisingly, one can find
at least one simple root. It is parametrized by the up to now free parameter $\mathfrak{p}_k$:
\be\label{expzk}
\mu_1 = \frac{\sin\left(\frac{p\mathfrak{p}_k}{p+1}\right)}{\sin\left(\frac{\mathfrak{p}_k}{p+1}\right)} e^{i\mathfrak{p}_k},
\quad z_k^{-1} = \frac{\sin^{p+1}(\mathfrak{p}_k)}{\sin\left(\frac{\mathfrak{p}_k}{p+1}\right) \sin^{p}\left(\frac{p\mathfrak{p}_k}{p+1}\right)}.
\ee
Then, using Vi\`ete's formulas \cite{Vinberg},
\be\label{Vieta}
e_i(\mu_1,\mu_2,\dots,\mu_{p+1}) = (-1)^i \left(\binom{p+1}{i}-z_k^{-1} \delta_{i,1}\right),
\ee
for $i=1,\dots,p+1$,
we can relate all the other roots to $\mathfrak{p}_k$. By means of (\ref{expzk}) the
quasi-particle energies in (\ref{EHgen})
can be written as,
\be\label{epansatz}
\epsilon_k =\frac{\sin^{\frac{p+1}{N}}(\mathfrak{p}_k)}{\sin^{\frac{1}{N}}
\left(\frac{\mathfrak{p}_k}{p+1}\right) \sin^{\frac{p}{N}}\left(\frac{p\mathfrak{p}_k}{p+1}\right)}\,,\quad k=1,\dots,\bar M\,.
\ee

Note that the second equation (\ref{eqAi}) fixes the amplitudes
$A_i$ in terms of the roots $\mu_i=\mu_i(\mathfrak{p}_k)$. The first equation in (\ref{eqAi})
then gives a quantization condition for $\mathfrak{p}_k$,
which, except for $p=1$, does not seem to have a closed solution. For $p=1$, one has,
\be
\mathfrak{p}_k = \frac{\pi}{\bar M+1}k,\quad k=1,\dots,\bar M\,,
\ee
such that the density of roots in the limit $\bar M\rightarrow \infty$ is
$\frac{\Delta \mathfrak{p}_k}{\Delta k}=\frac{\pi}{\bar M}$. Based
on numerical checks we claim that this result generalizes to arbitrary $p>1$,
including the case $p=2$ considered in \cite{Fendley:2019sdx}.

{\it\bf The ground-state energy per site in the bulk limit}.
The ground-state energy of the Hamiltonians given by (\ref{Hgen}) with
general $p$ are obtained by taking all the roots $\epsilon_k$ with $s_i=0$ in 
(\ref{EHgen}). Since the density of roots in the $\bar M\rightarrow \infty$ limit is,
\be
\frac{\Delta \mathfrak{p}_k}{\Delta k}=\frac{\pi}{\bar M}=\frac{(p+1)\pi}{M}\,,
\ee
we have,
\be\label{einf}
e_{\infty}^{(p)} \equiv -\frac{E_0}{M}= -\frac{1}{M}\sum_{k=1}^{\bar M}\epsilon_k=-\frac{1}{(p+1)\pi}
\int_{0}^\pi\epsilon(\mathfrak{p})d\mathfrak{p}\,.
\ee
By means of the change of variables $\sin\left(\frac{\mathfrak{p}_k}{p+1}\right)=\sin\left(\frac{\pi}{p+1}\right)\sqrt{t}$,
the integral (\ref{einf}) can be identified to an Euler integral in $(p-1)$ variables which is
the integral representation of the Lauricella hypergeometric series $F_D^{(p-1)}$ \cite{Slater}. The cases $p=1,2,3$
reduce to other known functions. For example, we have,
\be\label{ep1}
e_{\infty}^{(1)} = -\frac{2^{\frac{2}{N}-1}\Gamma\left(\frac{1}{N}+\frac{1}{2}\right)}{\sqrt{\pi}\left(\frac{1}{N}+1\right)}\,,
\ee
\be\label{ep2}
e_{\infty}^{(2)} = -\frac{3^{\frac{3}{N}+\frac{1}{2}}\Gamma\left(\frac{3}{N}+1\right)}
{2^{\frac{2}{N}+2}\sqrt{\pi}\Gamma\left(\frac{3}{N}+\frac{3}{2}\right)}
{}_2F_1\left(\begin{matrix}\frac{1}{2}\quad\frac{1}{N}+\frac{1}{2}\\ \frac{3}{N}+\frac{3}{2}\end{matrix} ;\frac{3}{4}\right)\,,
\ee
\be\label{ep3}
e_{\infty}^{(3)} &=& -\frac{2^{\frac{8}{N}-\frac{3}{2}}\Gamma\left(\frac{4}{N}+1\right)}
{3^{\frac{3}{N}}\sqrt{\pi}\Gamma\left(\frac{4}{N}+\frac{3}{2}\right)}
\non\\&\times&
F_1\left(\frac{1}{2};\frac{1}{2}-\frac{2}{N},\frac{3}{N};\frac{4}{N}+\frac{3}{2};\frac{1}{2},\frac{2}{3}\right)\,,
\ee
where $F_1$ is the Appel function,
\be\label{ep4}
&&e_{\infty}^{(4)} = -\frac{5^{\frac{5}{N}}\sin\left(\frac{\pi}{5}\right)\Gamma\left(\frac{5}{N}+1\right)}
{2^{\frac{8}{N}+1}\sqrt{\pi}\Gamma\left(\frac{5}{N}+\frac{3}{2}\right)}
\non\\&&\times
F_D^{(3)}\left(\frac{1}{2};\frac{1}{2}+\frac{2}{N},-\frac{5}{N},\frac{4}{N};\frac{5}{N}+\frac{3}{2};
x_1,x_2,x_3\right)\,
\ee
where $F_D^{(3)}$ is the Lauricella function with 3 variables at $x_1=\frac{1}{2+\frac{2}{\sqrt{5}}}$,
$x_2=\frac{2}{3+\sqrt{5}}$ and $x_3=\frac{1}{1+\frac{1}{\sqrt{5}}}$. We note that for $N=2$, the integrals converge to,
\be\label{einfn2}
e_{\infty}^{(p)} = -\frac{\Gamma\left(\frac{1}{2}+\frac{p}{2}\right)}{\Gamma\left(1+\frac{p}{2}\right)}\,\quad\textrm{for}\quad N=2.
\ee
The numerical solutions for the roots of (\ref{polP}) are simple to obtain numerically. In the third column of
Table \ref{tabgs} we present the values of $e_{\infty}^{(p)}$ obtained directly from the roots up to $\bar M=500$.
The results are the extrapolated ones. The agreement with the exact results obtained from
the integral (\ref{einf}), shown in the fourth column of Table \ref{tabgs}, is remarkable.
\begin{table}
\be
\begin{array}{c|c|c|c|c|c}
N & p &-e_{\infty}^{(p)}:~ \textrm{roots}  & -e_{\infty}^{(p)}:~ \textrm{exact} & z_{M}^{(p)}
& z_c\\ \hline 
2& 1 & 0.63661977 & 0.63661977 & 1.0000000 & 1 \\
2&2 & 0.50000000 & 0.50000000 & 1.4999998 & 3/2 \\
2 & 3 & 0.42441318  & 0.42441318& 1.9999994 & 2 \\
2 & 4 & 0.37500000 & 0.37500000& 2.4999996 & 5/2 \\\hline
3 &1 &  0.56604660 & 0.56604668 & 0.6666667 & 2/3 \\
3&2 & 0.41349667  & 0.41349667 & 0.9999999& 1 \\
3&3 & 0.33333331 & 0.33333333 & 1.3333331& 4/3 \\
3&4 & 0.28302332 & 0.28302334 & 1.6666663& 5/3 \\\hline
4&1 &  0.53935231 & 0.53935260 & 0.5000000& 1/2 \\
4&2 &  0.38137983 & 0.38137988 & 0.7499999& 3/4 \\
4&3 & 0.30010545 & 0.30010544 & 0.9999998 & 1 \\
4&4 & 0.25000000 & 0.25000000 &1.2499997& 5/4 \\\hline
\end{array}\non
\ee
\caption{Ground state energy per site $-e_{\infty}^{(p)}$ and dynamical critical exponent $z_c$ for the models (\ref{Hgen},\ref{H}), for some values
of $N$ and $p$. The third column are the extrapolated ($\bar M\rightarrow \infty$) results obtained from the roots of (\ref{polP}) up to $\bar M=500$. The exact results
(\ref{ep1},\ref{ep2},\ref{ep3},\ref{ep4}) are shown in the fourth column. The estimator $z_M^{(p)}$ (see text) and the exact results for the dynamical critical
exponent are shown in the last two columns.}
\label{tabgs}
\end{table}

The first excited state $E_1$ is obtained
by the addition of all the roots in \ref{EHgen} with  $s_i=0$ except for the largest one 
($s_{\bar M}=1$), where
$\mathfrak{p}_k=\mathfrak{p}_{\bar M}=\pi-\frac{a}{M}$, with $a$ being a harmless constant.
The real part of the energy gap behaves, as $M\rightarrow \infty$,
\be\label{gap}
\Delta_M^{(p)}=E_1-E_0=(1-\omega)\epsilon(\mathfrak{p}_{\bar M})
\approx
\left(\frac{a}{M}\right)^{z_c}\,,
\ee
where $z_c=(p+1)/N$ is the dynamical critical exponent. The case $N=2$ and
$p=1$ gives the dynamical exponent
of the conformally invariant quantum Ising chain ($z_c=1$), and the case $N=p=2$ recovers the result $z_c=3/2$
obtained in \cite{Fendley:2019sdx}.
For the sake of illustration, we also show in the fifth column of Table \ref{tabgs} the extrapolated results of the estimator $z_{M}^{(p)}=-\log(\Delta_M^{(p)})/\log(M)$
of the dynamical critical exponent for some values of $N$ and $p$. These results were obtained from
the numerical solution of the largest root of the polynomial (\ref{polP}). The agreement with the predicted result $z_c=(p+1)/N$ (last column)
is also remarkable.

The result (\ref{gap}) then imply that the Hamiltonians (\ref{Hgen}) with
the couplings (\ref{speccouplings}) and general values of $p$ undergoes a continuous phase
transition at $\lambda=\lambda_c=1$ with dynamical critical exponent $z_c=(p+1)/N$,
also shown in the last column of Table \ref{tabgs}.

{\it\bf Discussion}. We have introduced a new family of
integrable quantum spin chains with multispin interactions
and which have a free fermionic or parafermionic spectrum. We believe that
many interesting directions of investigation can be pursued.
Firstly, it would be important to prove the conjectures we have made, in particular
the product formula (\ref{prodformula}). Next, one should try to construct
the so-called parafermionic operators which satisfy a generalized Clifford algebra \cite{Fendley:2013snq,Baxter2014,Au_Yang_2014,auyang2016parafermions,Fendley:2019sdx}.
This is one step towards the computation of correlation functions for this family of models.
It is also very interesting to compute the entanglement entropy, initially for $N=2$ (in this case,
the Hamiltonian is hermitian) and arbitrary $p$, and study the role of the
polynomials (\ref{generalP}) along the lines of \cite{Crampe2019}. Another problem is to study the random coupling case.
We also think it is worth exploiting connections between the algebra (\ref{halgebra}) and the algebras
of Onsager and of Temperley-Lieb, as well as trying to consider the models within the quantum
inverse scattering method. Going beyond the free boundary case considered here, it is challenging to consider
the Hamiltonians with periodic boundary conditions; in this case, the product formula (\ref{prodformula}) is no longer valid, and
needs to be generalized.

\begin{acknowledgments}
RAP thanks S. Belliard and A. Morin-Duchesne for
discussions on the inversion relation for free fermionic systems.
The work of FCA was supported in part by the Brazilian agencies
FAPESP and CNPq. 
RAP was supported by FAPESP/CAPES (grant \# 2017/02987-8).
\end{acknowledgments}

\end{document}